\documentclass{article}

\usepackage{latexsym}
\usepackage{graphics}
\usepackage{amsmath}
\newcommand{\virgodate}{August 17, 2017}

\begin{document}

\noindent \huge \textbf{Using weighting algorithms to refine source direction determinations in all-sky gravitational wave burst searches with two-detector networks} \\
\normalsize
T. McClain \footnote {Correspondence to: tjamcclain2@gmail.com} \\
Weinberg Theory Group, University of Texas, Austin, Texas, USA \\

\renewcommand{\abstractname}{}
\begin{abstract}
\noindent I explore the possibility of resurrecting an old, non-Bayesian computational approach for inferring the source direction of a gravitational wave from the output of a two-detector network. The method gives the beam pattern response functions and time delay, and performs well even in the presence of noise and unexpected signal forms. I further suggest an improvement to this method in the form of a weighting algorithm that usefully improves its accuracy beyond what can be achieved with simple best-fit methods, validating the new procedure with several small-scale simulations. The approach is identified as complimentary to -- rather than in competition with -- the now-standard Bayesian approach typically used by the LIGO network in parameter determination. Finally, I briefly discuss the possible applications of this method in the world of three-or-more detector networks and some directions for future work.
\end{abstract}

\vspace{0.5in}


\section{Introduction}
\label{intro}
The use of modern computational methods to solve for the polarization amplitudes \( h_+(t) \) and \( h_{\times}(t) \) of a gravitational wave and to gain useful information about the direction to the source \( (\theta, \phi) \) goes back to the seminal paper by Gursel and Tinto \cite{gursel&tinto}. That paper details a method for extracting all four of these parameters from the noisy output of a three detector system in a method that is all-but-optimal and that requires no assumptions about the nature of the source. However, even though the three detector problem has been essentially solved since 1989, it was already clear at that time that problems emerged in the case of a two-detector network, and later analysis \cite{rakhmanov} confirmed that the two-detector problem poses a number of additional challenges and does not admit unique solutions. The actual method implemented by the LIGO collaboration to determine source parameters in their gravitational wave burst search is implemented in a code package called LALInference \cite{abbottLRR}; an explication of this method can be found in \cite{veitch}. Though the Bayesian inference framework of LALInference is the consensus of many teams and over a decade of research, the method as implemented in the first successful generation of LIGO science runs has at least three aspects upon which one might hope to improve: its reliance on model-specific waveforms, its high degree of final uncertainty, and its high computational cost. For example, the LALInference analysis of LIGO's early two detector data (VIRGO has since joined the network and participated in a joint detection on \virgodate) assumes that the source is one of only a few well-modeled astrophysical phenomena, reports sky-angles with uncertainties of \( 15-70 \% \) and average errors of \( 1-10 \% \), and takes on the order of months of CPU time to run \footnote{The possibility of parallelization means that CPU time is not the same as real time. However, many algorithms -- including LALInference -- make substantial use of techniques like MCMC that are notoriously resistant to parallelization. Nevertheless, the latest version of LALInference is actually run in about a day of real time \cite{singer2016rapid}.} \cite{veitch}. Recent improvements to the basic methods of LALInference have enabled more rapid characterization of gravitational wave transient detection candidates of the type considered in this paper \cite{lynch2017information} and much faster (on the order of minutes) determination of signal sky locations in some regions of the sky for some types of loosely-modeled sources \cite{klimenko2016method}. Meanwhile, a substantially different basic method, though still very much in the mode of Bayesian analysis, represents another way to very rapidly (again, within minutes) determine the sky locations of gravitational wave transients at relatively high signal to noise levels, independent of assumptions regarding the signal source \cite{singer2016rapid}. The results of this most recent generation of research show that rapid detection and sky localization is often (though not always) possible within the framework of Bayesian analysis. However, as pointed out in \cite{lynch2017information}, it is always worthwhile to have multiple algorithms available to cross-check and confirm results, especially if those algorithms differ enough in assumptions and implementation that they are unlikely to produce similar errors. It is in this spirit that I offer a rapid, non-Bayesian approach to sky location determination in gravitational wave detector networks.

The method to be discussed begins with a feature of the two detector system that was already known to Gursel and Tinto and is briefly discussed in \cite{gursel&tinto}: if the source angles \( (\theta, \phi) \) are known, the two detector problem is analytically solvable, and gives a unique value for each of \( h_+ \) and \( h_{\times} \) for each time at which the response of the two detectors is known. Unless the gravitational wave search is externally triggered (that is, induced by an optical astronomical observation), the sky angles are \emph{not} known in advance. However, if an algorithm could be successfully implemented that could find the sky angles -- without regard to any other model parameters -- then the problem would be analytically solvable with this additional input.

The most obvious matching algorithm to try is very much like the one on which LALInference is based: simulate a great many possible signals and look for the one with the smallest deviation from the actual detector responses. Even with a much-reduced model space (as compared to LALInference), this is method is workable, and can produce good results. However, it seems intuitively reasonable that, in the absence of any single excellent fit (as we must expect in the presence of substantial noise or if we do not wish to expend the extraordinary computational resources required to produce accurate results with LALInference), it should be the sky angles that produce the largest number of better-than-average fits that represent the true source location. This is the intuition underlying the new proposed fitting algorithms, and simulation results suggest that this method can not only reproduce the success of more standard fitting algorithms, but substantively improves upon the results of single-best-fit methods in almost every tested scenario.

\section{Methods}
\label{methods}

To establish the validity of this modified approach, it was first necessary to write a simplified implementation of the single-best-fit method underlying most other approaches. I began with the starting assumption of a sine-Gaussian waveform, as used by the LIGO collaboration in their all-sky burst search event detection algorithms \cite{abbott2009}. The success of a particular parameter set \( \Theta \) is quantified by summing the absolute value of the difference between the algorithm's calculated responses \( R_{out} \) and the simulated detector responses \(R_{in}\) for each detector in the network \( N \) and each sampled time  \( T \) in the lifetime of the signal: 
\begin{equation} 
Q(\Theta) := \sum_{n \in N \, , \, t \in T} \sqrt{ \left(R_{n,out}(t, \Theta) - R_{n,in}(t) \right)^2 }
\label{eq:quality}
\end{equation} 
The more common choice would be to normalize to the noise variance:
\begin{equation}
Q'(\Theta) := \sum_{n \in N \, , \, t \in T} \sqrt{ \left( \frac{(R_{n,out}(t, \Theta) - R_{n,in}(t))^2}{\eta^2_{n}(t)} \right) }
\end{equation}
where the term $\eta_n(t)$ represents the (estimated) noise in the $n^{th}$ detector at time $t$ Since this paper deals only with monochromatic signals, the noise is identical across all modeled parameter sets and there is no reason to normalize to the noise variance \footnote{Of course, this would not be the case if non-monochromatic signals were being analyzed after Fourier decomposition. In this case, we would normalize to the noise variance as usual.}. Since the algorithm is designed to sample the detector signal stream only very sparsely, one might expect poor results from allowing low-response pieces of the sample signal stream to be underweighted, as this effectively reduces the sampling rate even further. This intuition is not borne out by simulations over full signal lifespans, and eq. (\ref{eq:quality}) is used throughout the paper to determine the success of parameter sets.

The detector responses \( R_n(t) \) are calculated in the standard way (see, for example, \cite{schutz2011}) and can be made to include both randomly generated noise and other, non-random but un-modeled contributions to the waveform (that is, signal that does not conform to the sine-Gaussian model and therefore cannot be readily fit by the algorithm). Following the conventions used by Schutz in \cite{schutz2011}, we compute the response functions
\begin{equation}
R_{n} (t) = h_+(t + \tau_n) F^+_n (\theta, \phi) + h_{\times} (t + \tau_n) F^{\times}_n (\theta, \phi) +\eta_n(t) 
\end{equation}
where \( h_+ \) and \( h_{\times} \) represent the two independent polarization amplitudes of the incoming gravitational wave, \( \tau_n = \frac{1}{c} (\vec r_0 - \vec r_n) \cdot \hat e_{gw} \) represents the time delay between the \( n^{th} \) detector and an arbitrarily chosen `` \( 0^{th} \)'' reference detector, \( F^+_n \) and \( F^{\times}_n \) represent the beam pattern response functions of the \( n^{th} \) detector (that is, the response of the \( n^{th} \) detector to a unit-amplitude, linearly polarized signal \( h_+ = 1 \) or \( h_{\times} = 1 \) ), and \( \eta_n \) represents the noise in the \( n^{th} \) detector. Because actual gravitational wave detectors seem to have instrument noise that does not necessarily match the Gaussian noise model (see, for example, \cite{abbott2016characterization}), this analysis assumes non-Gaussian noise. This noise is random-number generated, and is characterized throughout the paper by its maximum allowed value within a given set of simulations, \( \eta_{max} \), which is in turn set by the signal-to-noise ratio chosen for each simulation set: \( \frac{\sqrt{h^2_{+,max} + h_{\times, max}^2}}{\eta_{max}} = \text{SNR} \). The noise values in each detector are generated independently, and each is uniformly distributed within the range \( [ - \eta_{max}, \eta_{max} ] \).

For the purposes of this algorithm, the beam pattern response functions are fully general (see, for example, \cite{schutz2011} or \cite{anderson2001}, though I follow different angle conventions than the latter source). I modeled only monochromatic, sine-Gaussian signals of the form

\begin{equation}
s_{+ \, , \, \times} (t, q, \omega, a_n) = \exp (- q^2 t^2) (a_1 \cos \omega t + a_2 \sin \omega t)
\end{equation}
though with the potential for the ``real'' (simulated) signal \( h_+ \, , \, h_{\times} \) to be modified by an arbitrary, un-modeled function, which the algorithm includes up to fifth order in \( t \) :
\begin{equation}
h_{+ \, , \, \times} (t, q, \omega, a_n, u_n) = \exp (- q^2 t^2) (a_1 \cos \omega t + a_2 \sin \omega t) (1+u_1 t + u_2 t^2 + u_3 t^3 + u_4 t^4 + u_5 t^5)\
\label{eq:unmodeledsignal}
\end{equation}
As with the noise, these un-modeled signal amplitudes are random-number generated and are characterized throughout the paper by their maximum allowed values within a given set of simulations. The fitting algorithms can easily be made to handle non-monochromatic signals after Fourier decomposition at the expense of greater computational cost; I have avoided these extra computational costs in this analysis. 

Though my current modeling does not replicate much of the current state-of-the-art used, for example, in LALInference, the model does contain a few of the interesting features of LALInference, for example, quasi-random sampling of the parameter space \cite{veitch}. It uses a more computationally intensive method of determining best fit \cite{sathyaprakash}. More importantly, it is also designed to weigh many different fits with non-minimal \( Q \) values in the final determination of the ``best fit'' sky angles. Specifically, it allows searches other than the single best fit found by maximizing the weighting function 

\begin{equation}
W = 
\begin{cases}
1 & \mbox{if} \ Q = Q_{min} \\
0 & \mbox{otherwise} \\
\end{cases} 
\label{eq:minQ}
\end{equation}
This weighting algorithm simply looks for the parameter set with the minimum value of \( Q \). My algorithm instead produces a weighted best fit after summing over all parameters with which we are not concerned by minimizing an arbitrary weighting function \( W = f (Q/Q_{min}) \)
\begin{equation}
\min \left( \sum_{\Theta \setminus (\theta , \phi)} f \left( \frac{Q(\Theta)}{Q_{min}} \right) \right)
\end{equation}
The output of the algorithm consists of the \( F^+ \), \( F^{\times} \), and \( \tau \) parameters that result from the \( (\theta, \phi ) \) values of the parameter set that minimizes this weighted sum.

It is worth pointing out explicitly that we do not need the sky angles \( \theta \) and \(\phi \) if our only goal is to reconstruct \( h_+ \) and \( h_{\times} \). Rather, it is \( F^+_n(\theta, \phi) \), \( F^{\times}_n(\theta, \phi) \), and \( \tau_n(\theta, \phi) \) that determine whether a particular set of sky angles accurately determines the response functions measured in the detector network. Any sky angles that produce the same values of \( F^+_n(\theta, \phi) \) and \( F^{\times}_n(\theta, \phi) \) and the same time delays \( \tau_n(\theta, \phi) \) will produce the same detector responses. Since the beam pattern response functions are multi-valued, there is typically not a single pair of sky angles determined by the algorithm. Consequently, for a two detector network what the algorithm most accurately determines are the beam pattern response functions of the two sites ( \( F^+_1, F^{\times}_1, F^+_2, F^{\times}_2 \) ), as well as the time delay \( \tau \) between the two sites. Again, it is this information that is needed to find \( h_+, h_{\times} \) analytically (and again, the details can be found in \cite{gursel&tinto}). The reason the algorithm computes all relevant parameters in terms of sky angles is to ensure that all modeled results are physically realizable. The actual values of these sky angles \( (\theta, \phi) \) are not necessary to solve the problem of determining \( h_+ \) and \( h_{\times} \), but they can be recovered (non-uniquely) by finding the intersection of the multi-valued inverses of the beam pattern and time delay functions.

\section{Results}
\label{results}

For this analysis, I characterize the performance of the various algorithms by the RMS difference between the model's predicted values for \( F^+ \) and \( F^{\times} \) at each site, the model's predicted \( \tau \) between the two sites, and the actual values of those five functions calculated from the randomly generated values \( ( \theta, \phi ) \) that produce the simulated signal. Values with \( in \) subscripts denote the values that serve as the algorithm's (simulated) input values (in place of real signal data), and \( out \) subscripts denote the values predicted by the algorithm:
\begin{equation}
\delta F_{rms} :=  \frac{1}{4} \sqrt{(F^+_{1,out} - F^+_{1,in})^2 + (F^{\times}_{1,out} - F^{\times}_{1,in})^2 + (F^+_{2,out} - F^+_{2,in})^2 + (F^{\times}_{2,out} - F^{\times}_{2,in})^2}
\end{equation}
\begin{equation}
\delta \tau_{rms} := \sqrt{ (\tau_{out} - \tau_{in})^2}
\end{equation}
For specificity, I have chosen the two currently operating LIGO network detectors as the sites, with the Hanford detector set to be the reference site (\( \tau = 0 \)).

To allow a modestly fine sampling of the angular parameter space, I first (and primarily) report the results of the highly specialized situation in which the frequency and q-factor of the incoming gravitational wave are known in advance, and the wave is assumed to be circularly polarized (see, for example, \cite{aasi2015} for more generic conditions, as well as below). This could, however, be a reasonable approximation to situations in which the inclination of the source system is known to be almost face-on, the signal is Fourier decomposed, and the analysis is then performed on a handful of dominant frequencies. The parameter set used is in these simulations is: \( t \in \left[ 0, 0.19 \right] \, \text{s} \), \(q = 4.3 \, \text{s}^{-1} \), \( f \in \{10, 100, 1000 \} \, \text{Hz} \), \( \theta \in \left[ 0, \pi \right] \, \text{rad} \), \( \phi \in \left[ 0, 2 \pi \right] \, \text{rad} \), circular polarization, \( \text{SNR} \in \left[ 3, 100 \right] \), and the parameters \( u_1 \) through \( u_5 \) of eq. (\ref{eq:unmodeledsignal}) all capped at a single value \( u_{\text{max}} \in \left[ 1/100, 1/3 \right] \). Each simulation is run with different random values for the input parameters, as well as freshly randomized noise and un-modeled signal, all within the appropriate bounds. Few times are checked over the lifetime of the signal due to computational restrictions. 

Naturally, the performance of a revised algorithm with weighting function \( f (Q/Q_{min}) \) depends entirely upon the function \( f \). Extensive but non-exhaustive testing with weighting functions that are logarithmic, polynomial, and exponential in the argument \( Q/Q_{min} \) ultimately resulted in my choice to focus on weighting functions of the form
\begin{equation} 
W = \exp \left[ 1 - \left( \frac{Q}{Q_{min}} \right)^n \right]
\label{eq:weightedQ}
\end{equation}
for values of \(n \in \left[ \frac{1}{2} , 64 \right] \). These weighting functions peak at \( Q = Q_{\min} \) with a value of \( 1 \), and reduce to exactly the result of eq. (\ref{eq:minQ}) in the limit \( n \rightarrow \infty \). More intuitively, these functions look ``almost'' like the weighting function of eq. (\ref{eq:minQ}), but with some non-zero range over which ``good'' fits that are non-minimal can still contribute to the determination of the final angle values. For \( n = 2 \), the weighting function of eq. (\ref{eq:weightedQ}) is a Gaussian centered on the minimum \( Q \) value.

As expected, the weighting functions of eqs. (\ref{eq:weightedQ}) and (\ref{eq:minQ}) both perform substantially better than a ``random choice'' weighting function that simply weights every possible parameter value \( ( \theta, \phi ) \) the same, even at very high noise levels and in the presence of substantial un-modeled signal. At every tested noise level there is one or more weighting function of the form of eq. (\ref{eq:weightedQ}) that substantially improves upon the performance of the weighting function of eq. (\ref{eq:minQ}). As noise decreases, accuracy increases for the weighting functions of eq. (\ref{eq:minQ}) and eq. (\ref{eq:weightedQ}).

\begin{figure*}
\centering
\includegraphics{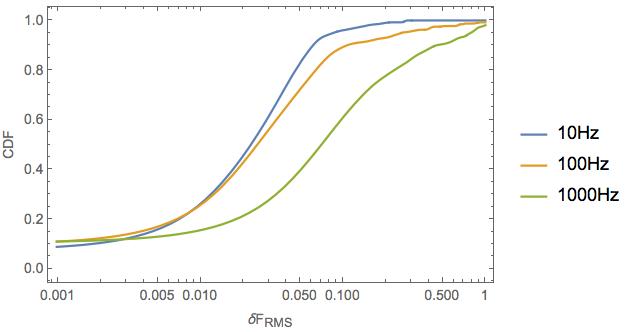} 
\caption{Cumulative probability distribution of $\delta F_{rms}$ for the value $n=4$ at each tested frequency.}
\label{fig:freqsplot}
\end{figure*}

\begin{figure*}
\centering
\includegraphics{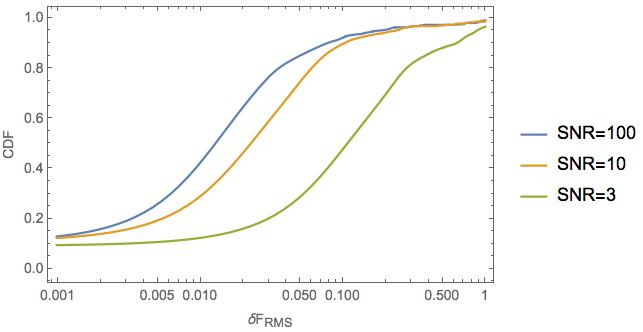} 
\caption{Cumulative probability distribution of $\delta F_{rms}$ for the values $\{ n=2 \, , \, n=4 \, , \, n=2 \}$ for the given maximum values of noise and un-modeled signal.}
\label{fig:snrsplot}
\end{figure*}

To give a sense of the raw accuracy of the method, figs. \ref{fig:freqsplot} and \ref{fig:snrsplot} show the cumulative probability distribution function of $\delta F_{rms}$ over the frequencies and a few of the (maximum) SNRs and un-modeled signal fractions tested, respectively. As expected, higher frequency signals and higher maximum levels of noise and un-modeled signal substantially reduce the accuracy of the method, but we note that the method does not completely lose predictive power even at high frequencies and high noise/un-modeled signal.

To give a sense of how the algorithm might be deployed in a more realistic gravitational wave burst search, I have done simulations that drop the assumption of circular polarization, allowing an arbitrary sine-Gaussian waveform while maintaining as much as possible the fineness of the parameter space mesh. To compensate for the additional computational load associated with the many extra degrees of freedom in this parameter space, I drastically reduced the size of the angular search space: the reduced parameter angular parameter spaces are randomly distributed throughout the sky with angular windows of $\pi/20 \, \text{rad}$ for both $\theta$ and $\phi$. Though it seems artificial, this reduction is consistent with the most natural way in which the algorithm is parallelizable: by supplying many cores with the same parameter space model but focusing each on a small subspace of the angular part of that parameter space -- in principle all the way down to a single pair of angles $ ( \theta, \phi ) $ -- it should be possible to vastly decrease the computational load on each core. However, doing this as I have without the rest of the angular parameter space being evaluated on other cores means that the raw accuracy of this method is not being tested; indeed, even just taking the average $F$ values associated with this drastically reduced angular space results in a reasonably good fit. However, it does allow us to probe the relative capabilities of the weighting functions of eq. (\ref{eq:minQ}) and eq. (\ref{eq:weightedQ}). As anticipated, the same values of $n$ for which the weighting functions of eq. (\ref{eq:weightedQ}) seem to outperform those of eq. (\ref{eq:minQ}) in the case of circular polarization are also superior in the case of elliptical polarization. For example, at $f = 100 \text{Hz}$, with independent noises $\eta_n$ and each order of un-modeled signal \( u_1 \) through \( u_5 \) of eq. (\ref{eq:unmodeledsignal}) capped at $1/10$ of the maximum signal value \( \sqrt{h_{+, max}^2 + h_{\times, max}^2} \), $n=4$ outperforms the single best fit result by $3.2 \%$ in $\delta F_{rms}$ and $17.3 \%$ in $\delta \tau_{rms}$ in preliminary simulations. Though much more extensive testing with a set of networked cores is necessary to prove that the algorithm will continue to perform well once massively parallelized, the current simulations results seem reasonably promising. 

\section{Discussion and conclusions}
\label{discussion}

The simulations run thus far indicate that the sharply-decreasing exponential weighting of eq. (\ref{eq:weightedQ}) outperforms the single best fit weighting of eq. (\ref{eq:minQ}) in many scenarios. Most importantly, my results indicate that the revised weighting is appreciably superior to the the min-Q weighting method in the presence of substantial noise and/or un-modeled signal; i.e., in precisely the scenario that most closely matches the actual all-sky burst search of the LIGO network. Indeed, we find that the weighting functions of eq. (\ref{eq:weightedQ}) with \( n \in \{ 2, 4 \} \) outperform the standard algorithm of eq. (\ref{eq:minQ}) at almost every noise level tested. This preliminary analysis indicates that use of a revised weighting algorithm of the form of eq. (\ref{eq:weightedQ}) is likely to produce a substantial improvement over the algorithm of eq. (\ref{eq:minQ}) even in situations in which the noise in the signal cannot be accurately quantified. Even better results can be obtained if the amount of noise can be quantified in advance of applying the weighting algorithm and the power \( n \) in the weighting function chosen appropriately; see figs. \ref{fig:freqsplot} and \ref{fig:snrsplot} for two examples. If the amount of noise cannot be quantified prior to the analysis, \( n = 4 \) provides the best average result at the noise levels tested. To reiterate, the revised weighting algorithm can typically provide a substantial improvement in performance over the standard algorithm of eq. (\ref{eq:minQ}) regardless of whether or not the noise level is quantified before the analysis.

To achieve this improved output in the context of an algorithm like LALInference, it would be necessary to use a finer mesh of parameter values, but the computational cost of doing so is far greater than that of implementing the revised weighting algorithm. To establish a baseline, we first consider the cost of obtaining a particular result with the weighting function of eq. (\ref{eq:minQ}). Using this weighting function does not require us to calculate the large array of weighted \( Q \) values that we must have if we wish to use the weighting function of eq. (\ref{eq:weightedQ}); the computational cost of calculating this large array represents the additional computational load of the revised weighting algorithm. In the case of the monochromatic sine-Gaussian model (much simplified compared to LALInference, with only eight parameters), using the $n=2$ weighting algorithm on a signal of frequency $f = 100 \text{Hz}$ with noise and un-modeled signal each less than $1/10$ of the signal value is expected to produce about a \( 13 \% \) average improvement in accuracy based on the results of current simulations \footnote{We use the circular polarization results for making this prediction because this scenario has been tested much more extensively. The relatively few simulations so far carried out with elliptically polarized signals seem to indicate that the elliptical results will mirror the circular ones.}, while increasing the computational load of the model by \( 5.0 \% \) based on direct measurement of simulation runtimes. On the other hand, to achieve this same average improvement in accuracy using a finer mesh of parameters values requires using a \( 13 \% \) finer mesh, which in turn increases the computational load of the model by \( 1.13^8 \approx 250 \% \): every large array in the simulation grows by that same factor, and the computational cost of the simulation is dominated by a few computations involving those large arrays. The recommended strategy would therefore be to use the finest parameter mesh possible with a given set of computational resources, and then to implement the revised weighting algorithm to achieve a final \( 13 \% \) improvement in accuracy without much attendant increase in computational load.

The ideal algorithm would excel in all three areas where the current LALInference method could potentially be improved: its number of model-specific assumptions from computational general relativity, the uncertainty and error of the final fit parameters, and its computational cost. The method proposed in this paper avoids any specific input from numerical general relativity, shows no sharp threshold below which it loses predictive power (as seems to be common with Bayesian methods; see, for example, \cite{singer2016rapid} and \cite{lynch2017information}), and promises to be parallelizable to run in as little as $10 - 20 \text{s}$ on ordinary desktop computers, it appears to make some progress in each of these areas. The present analysis does not prove that this algorithm can be made to reach the same degree of uncertainty or error attainable by LALInference. However, the apparent success of this new algorithm in several areas of interest would seem to indicate that it may be worthwhile to spend the computational resources necessary to determine whether the new weighting algorithm can be made fully competitive in this regard.

The careful reader will have no doubt noticed that no new method has been suggested for using this algorithm's output (the beam pattern response functions \( F^+ \) and \( F^{\times} \) at the two sites, as well as the time delay \( \tau \) between the sites) to solve for the polarization amplitudes \( h_+ \) and \( h_{\times} \). As alluded to in the introduction, this problem has already been solved \cite{gursel&tinto}. The non-algorithmic novelty of this method consists in a change of perspective: we seek only to establish beam pattern response functions numerically, and then solve for the polarization amplitudes algebraically, rather than trying to solve for these parameters (or, indeed, all relevant model parameters, as in the case of LALInference) numerically.

Even if this method is successfully implemented with uncertainty and error comparable to that of LALInference (while maintaining its relative stability under high noise and un-modeled signal and its relatively low computational cost), it is not designed to provide the other model parameters that can - in principle - be determined by LALInference. This inability is, of course, intimately related to one of the algorithm's major strengths, namely its lack of model-specific assumptions: without assuming a particular numerical model for the incoming signal, one cannot expect to extract model-specific parameter values for the source. 

As a result, this method and algorithm should not be viewed as standing in competition with LALInference (or other, similar, modeling efforts). Its best possible use would be to supplement more model-specific and computationally expensive methods. For example, when a small enough range of true sky angle \( (\theta, \phi ) \) pairs can be extracted from this algorithm's output, its relatively low computational cost (once parallelized) could allow it to be used as a cross-check for immediate (real-time) follow up to a detected signal using optical astronomy, something that has only recently become possible using Bayesian algorithms in the tradition of LALInference. Another possibility is that the algorithm can be used to rapidly characterize \( ( h_+ \, , \, h_{\times} ) \) so that other (non-Bayesian) computational methods lying outside the scope of LALInference can be brought to bear on these intrinsic gravitational wave parameters as soon as possible after a signal is detected, then independently compared to the output of LALInference (and its rapid sky angle determination partners) once it is available. More detailed simulations will show whether this algorithm can achieve the low levels of uncertainty and error necessary to support these potential uses.

As VIRGO began participating in the LIGO network on August 1, 2017 and took part in a joint detection of GW170814 on August 14, 2017 \cite{abbott2017gw170814}, we are finally working with networks of three detectors, with more anticipated to come on-line in the future. Since the three-detector problem is analytically solvable (as detailed in \cite{gursel&tinto}), one may question the utility of an approach specific to two-detector networks. The first point to make is that there may be times in the future when only a subset of detectors within a larger network actually makes a detection. This may happen, for example, due to one or more detectors being offline, or only a subset of the detectors having the sensitivity to detect the signal. It is estimated  that all three detectors will be simultaneously online only $50\%$ of the time (see, for example, \cite{singer2016rapid}), so this is by no means an unrealistic scenario. In this case, the two-detector situation may again become directly relevant. However, even in the case that three or more detectors all participate, the presence of noise makes analytical techniques potentially unreliable, and there may be benefit to a numerical approach like the one outlined in this paper. This could be applied in several possible ways, either in a coherent approach (in the style of \cite{klimenko2008coherent}) or by applying the method separately to each pair in the network and looking for coincidences (in the style of \cite{klimenko2004performance}). Future work will show whether this algorithm can support one or more of these potential applications.

\bibliographystyle{unsrt}
\bibliography{gw.bib}

\end{document}